\documentclass[twoside,11pt]{article}

\usepackage{jmlr2e}
\usepackage{amsmath}
\usepackage{algorithm}
\usepackage{algorithmic}
\usepackage{subfigure}
\usepackage{graphicx}
\usepackage{bbm}
\usepackage{tikz}
\usetikzlibrary{arrows}
\usetikzlibrary{graphs}
\usepackage{url}

\ShortHeadings{IFM LAB TUTORIAL SERIES \# 6, COPYRIGHT \copyright IFM LAB}{JIAWEI ZHANG, IFM LAB DIRECTOR}
\firstpageno{1}

\begin{document}

\title{Cognitive Functions of the Brain: Perception, Attention and Memory}

\author{\name Jiawei Zhang \email jiawei@ifmlab.org \\
	\addr{Founder and Director}\\
       {Information Fusion and Mining Laboratory}\\
       (First Version: May 2019; Revision: May 2019.)}

\maketitle

\begin{abstract}

This is a follow-up tutorial article of \cite{zhang2019secrets} and \cite{zhang2019basic}, in this paper, we will introduce several important cognitive functions of the brain. Brain cognitive functions are the mental processes that allow us to receive, select, store, transform, develop, and recover information that we've received from external stimuli. This process allows us to understand and to relate to the world more effectively. Cognitive functions are brain-based skills we need to carry out any task from the simplest to the most complex. They are related with the mechanisms of how we learn, remember, problem-solve, and pay attention, etc. To be more specific, in this paper, we will talk about the perception, attention and memory functions of the human brain. Several other brain cognitive functions, e.g., arousal, decision making, natural language, motor coordination, planning, problem solving and thinking, will be added to this paper in the later versions, respectively. Many of the materials used in this paper are from wikipedia and several other neuroscience introductory articles, which will be properly cited in this paper. This is the last of the three tutorial articles about the brain. The readers are suggested to read this paper after the previous two tutorial articles on brain structure and functions \cite{zhang2019secrets} as well as the brain basic neural units \cite{zhang2019basic}.

\end{abstract}


\tableofcontents

%--------------------------------------------------------------------------------------------------------------------
%--------------------------------------------------------------------------------------------------------------------
%--------------------------------------------------------------------------------------------------------------------

\section{Introduction}

As described in \cite{cognition}, cognition is the mental action or process of acquiring knowledge and understanding through thought, experience, and the senses. Human cognition can be conscious and unconscious, concrete or abstract, as well as intuitive (like knowledge of a language) and conceptual (like a model of a language). It encompasses many aspects of intellectual functions and processes such as attention, the formation of knowledge, memory and working memory, judgment and evaluation, reasoning and computation, problem solving and decision making, comprehension and production of language. Traditionally, emotion was not thought of as a cognitive process, but now much research is being undertaken to examine the cognitive psychology of emotion; research is also focused on one's awareness of one's own strategies and methods of cognition, which is called metacognition. Cognitive processes use existing knowledge and generate new knowledge.

%----------------------------------------------------------
\begin{table}[t]
\vspace{-1em}
\caption{Cognitive Development in Children.}
\centering
\begin{tabular}{| p{2.5cm} | p{3cm} | p{9cm} |}
\hline
\textbf{Stage} & \textbf{Age or Period} & \textbf{Description} \\
\hline\hline
Sensorimotor stage	&Infancy (0-2 years)	&Intelligence is present; motor activity but no symbols; knowledge is developing yet limited; knowledge is based on experiences/ interactions; mobility allows child to learn new things; some language skills are developed at the end of this stage. The goal is to develop object permanence; achieves basic understanding of causality, time, and space.\\
\hline
Pre-operational stage	&Toddler and Early Childhood (2-7 years)	&Symbols or language skills are present; memory and imagination are developed; nonreversible and nonlogical thinking; shows intuitive problem solving; begins to see relationships; grasps concept of conservation of numbers; egocentric thinking predominates.\\
\hline
Concrete operational stage	&Elementary and Early Adolescence (7-12 years)	&Logical and systematic form of intelligence; manipulation of symbols related to concrete objects; thinking is now characterized by reversibility and the ability to take the role of another; grasps concepts of the conservation of mass, length, weight, and volume; operational thinking predominates nonreversible and egocentric thinking.\\
\hline
Formal operational stage	&Adolescence and Adulthood (12 years and on)	&Logical use of symbols related to abstract concepts; Acquires flexibility in thinking as well as the capacities for abstract thinking and mental hypothesis testing; can consider possible alternatives in complex reasoning and problem solving.\\
\hline
\end{tabular}
\label{table:cognitive_development}
\vspace{-1em}
\end{table}
%----------------------------------------------------------

Jean Piaget was one of the most important and influential people in the field of ``developmental psychology''. He believed that humans are unique in comparison to animals because we have the capacity to do ``abstract symbolic reasoning''. His work can be compared to Lev Vygotsky, Sigmund Freud, and Erik Erikson who were also great contributors in the field of ``developmental psychology''. Today, Piaget is known for studying the cognitive development in children. He studied his own three children and their intellectual development and came up with a theory that describes the stages children pass through during development. The cognitive development at different stages in children is also illustrated in Table~\ref{table:cognitive_development}.

%----------------------------------------------------------
\begin{table}[t]
\vspace{-1em}
\caption{Cognitive Development in Children.}
\centering
\begin{tabular}{| p{4cm} | p{11cm} |}
\hline
 \textbf{Cognitive Ability} & \textbf{Detailed Description} \\
\hline\hline
 \textbf{Perception} &Recognition and interpretation of sensory stimuli (smell, touch, hearing, etc.)\\
\hline
 \textbf{Attention} & Ability to sustain concentration on a particular object, action, or thought, and ability to manage competing demands in our environment.\\
\hline
 \textbf{Memory} & Short-term/working memory (limited storage), and Long-term memory (unlimited storage).\\
\hline
 \textbf{Motor Skills} & Ability to mobilize our muscles and bodies, and ability to manipulate objects.\\
\hline
 \textbf{Language} & Skills allowing us to translate sounds into words and generate verbal output.\\
\hline
 \textbf{Visual and Spatial Processing} & Ability to process incoming visual stimuli, to understand spatial relationship between objects, and to visualize images and scenarios.\\
\hline
 \textbf{Executive Functions} & Abilities that enable goal-oriented behavior, such as the ability to plan, and execute a goal. These include:
\begin{itemize}
\vspace{-0.5em}
\item \textbf{Flexibility}: the capacity for quickly switching to the appropriate mental mode.
\vspace{-0.5em}
\item \textbf{Theory of mind}: insight into other people's inner world, their plans, their likes and dislikes.
\vspace{-0.5em}
\item \textbf{Anticipation}: prediction based on pattern recognition.
\vspace{-0.5em}
\item \textbf{Problem-solving}: defining the problem in the right way to then generate solutions and pick the right one.
\vspace{-0.5em}
\item \textbf{Decision making}: the ability to make decisions based on problem-solving, on incomplete information and on emotions (ours and others').
\vspace{-0.5em}
\item \textbf{Emotional self-regulation}: the ability to identify and manage one's own emotions for good performance.
\vspace{-0.5em}
\item \textbf{Sequencing}: the ability to break down complex actions into manageable units and prioritize them in the right order.
\vspace{-0.5em}
\item \textbf{Inhibition}: the ability to withstand distraction, and internal urges.
\vspace{-0.5em}
\end{itemize}\\
\hline
\end{tabular}
\label{table:cognitive_functions}
\vspace{-1em}
\end{table}
%----------------------------------------------------------

While few people would deny that cognitive processes are a function of the brain, a cognitive theory will not necessarily make reference to the brain or to biological processes. It may purely describe behavior in terms of information flow or function. Relatively recent fields of study such as neuropsychology aim to bridge this gap, using cognitive paradigms to understand how the brain implements the information-processing functions, or to understand how pure information-processing systems (e.g., computers) can simulate human cognition.

According to \cite{cognition_table}, several important cognitive functions of the brain, but not limited to, are briefly described in Table~\ref{table:cognitive_functions}, which include \textit{perception}, \textit{attention}, \textit{memory}, \textit{motor skills}, \textit{language}, \textit{visual and spatial processing} and \textit{executive functions}. In the following sections of this paper, we will provide more detailed descriptions about several main cognitive functions, including: perception, attention and memory, respectively.

%--------------------------------------------------------------------------------------------------------------------
%--------------------------------------------------------------------------------------------------------------------
%--------------------------------------------------------------------------------------------------------------------

\section{Perception}

As introduced in \cite{perception}, perception is the organization, identification, and interpretation of sensory information in order to represent and understand the presented information, or the environment. All perception involves signals that go through the nervous system, which in turn result from physical or chemical stimulation of the sensory system. For example, vision involves light striking the retina of the eye, smell is mediated by odor molecules, and hearing involves pressure waves. Perception is not only the passive receipt of these signals, but it's also shaped by the recipient's learning, memory, expectation, and attention. Generally, perception can be split into two processes:
\begin{itemize}
\item processing the sensory input, which transforms these low-level information to higher-level information (e.g., extracts shapes for object recognition);
\item processing which is connected with a person's concepts and expectations (or knowledge), restorative and selective mechanisms (such as attention) that influence perception.
\end{itemize}

Perception depends on complex functions of the nervous system, but subjectively seems mostly effortless because this processing happens outside conscious awareness. The perceptual systems of the brain enable individuals to see the world around them as stable, even though the sensory information is typically incomplete and rapidly varying. Human and animal brains are structured in a modular way, with different areas processing different kinds of sensory information. Some of these modules take the form of sensory maps, mapping some aspect of the world across part of the brain's surface. These different modules are interconnected and influence each other. For instance, taste is strongly influenced by smell.

Human perception abilities are heavily dependent on the brain as well as the surrounding sensory systems as introduced in \cite{zhang2019secrets}. The readers are suggested to refer to \cite{zhang2019secrets}, especially the Section 3, when reading the following contents.

\subsection{Detailed Process of Perception}

According to \cite{perception}, the process of perception begins with an object in the real world, termed the distal stimulus or distal object. By means of light, sound or another physical process, the object stimulates the body's sensory organs. These sensory organs transform the input energy into neural activity-a process called transduction. This raw pattern of neural activity is called the proximal stimulus. These neural signals are transmitted to the brain and processed. The resulting mental re-creation of the distal stimulus is the percept.

An example would be a shoe. The shoe itself is the distal stimulus. When light from the shoe enters a person's eye and stimulates the retina, that stimulation is the proximal stimulus. The image of the shoe reconstructed by the brain of the person is the percept. Another example would be a telephone ringing. The ringing of the telephone is the distal stimulus. The sound stimulating a person's auditory receptors is the proximal stimulus, and the brain's interpretation of this as the ringing of a telephone is the percept. The different kinds of sensation such as warmth, sound, and taste are called sensory modalities.

Psychologist Jerome Bruner has developed a model of perception. According to him, people go through the following process to form opinions:
\begin{itemize}
\item When we encounter an unfamiliar target, we are open to different informational cues and want to learn more about the target.
\item In the second step, we try to collect more information about the target. Gradually, we encounter some familiar cues which help us categorize the target.
\item At this stage, the cues become less open and selective. We try to search for more cues that confirm the categorization of the target. We also actively ignore and even distort cues that violate our initial perceptions. Our perception becomes more selective and we finally paint a consistent picture of the target.
\end{itemize}

According to Alan Saks and Gary Johns, there are three components to perception.
\begin{itemize}
\item The Perceiver, the person who becomes aware about something and comes to a final understanding. There are 3 factors that can influence his or her perceptions: experience, motivational state and finally emotional state. In different motivational or emotional states, the perceiver will react to or perceive something in different ways. Also in different situations he or she might employ a ``perceptual defence'' where they tend to ``see what they want to see''.
\item The Target. This is the person who is being perceived or judged. ``Ambiguity or lack of information about a target leads to a greater need for interpretation and addition.''
\item The Situation also greatly influences perceptions because different situations may call for additional information about the target.
\end{itemize}

Stimuli are not necessarily translated into a percept and rarely does a single stimulus translate into a percept. An ambiguous stimulus may be translated into multiple percepts, experienced randomly, one at a time, in what is called multistable perception. And the same stimuli, or absence of them, may result in different percepts depending on subject's culture and previous experiences. Ambiguous figures demonstrate that a single stimulus can result in more than one percept; for example the Rubin vase which can be interpreted either as a vase or as two faces. The percept can bind sensations from multiple senses into a whole. A picture of a talking person on a television screen, for example, is bound to the sound of speech from speakers to form a percept of a talking person.

\subsection{Types of Perception}

As introduced in \cite{perception}, human brain can sense different types of perceptions, and we will introduce several important perception types in this section as follows.
\begin{itemize}

\item \textbf{Vision}: In many ways, vision is the primary human sense. Light is taken in through each eye and focused in a way which sorts it on the retina according to direction of origin. A dense surface of photosensitive cells, including rods, cones, and intrinsically photosensitive retinal ganglion cells captures information about the intensity, color, and position of incoming light. Some processing of texture and movement occurs within the neurons on the retina before the information is sent to the brain. In total, about 15 differing types of information are then forwarded to the brain proper via the optic nerve.

\item \textbf{Sound}: Hearing (or audition) is the ability to perceive sound by detecting vibrations. Frequencies capable of being heard by humans are called audio or sonic. The range is typically considered to be between 20 Hz and 20,000 Hz. Frequencies higher than audio are referred to as ultrasonic, while frequencies below audio are referred to as infrasonic. The auditory system includes the outer ears which collect and filter sound waves, the middle ear for transforming the sound pressure (impedance matching), and the inner ear which produces neural signals in response to the sound. By the ascending auditory pathway these are led to the primary auditory cortex within the temporal lobe of the human brain, which is where the auditory information arrives in the cerebral cortex and is further processed there.

Sound does not usually come from a single source: in real situations, sounds from multiple sources and directions are superimposed as they arrive at the ears. Hearing involves the computationally complex task of separating out the sources of interest, often estimating their distance and direction as well as identifying them.

\item \textbf{Touch}: Haptic perception is the process of recognizing objects through touch. It involves a combination of somatosensory perception of patterns on the skin surface (e.g., edges, curvature, and texture) and proprioception of hand position and conformation. People can rapidly and accurately identify three-dimensional objects by touch. This involves exploratory procedures, such as moving the fingers over the outer surface of the object or holding the entire object in the hand. Haptic perception relies on the forces experienced during touch.

Gibson defined the haptic system as ``The sensibility of the individual to the world adjacent to his body by use of his body''. Gibson and others emphasized the close link between haptic perception and body movement: haptic perception is active exploration. The concept of haptic perception is related to the concept of extended physiological proprioception according to which, when using a tool such as a stick, perceptual experience is transparently transferred to the end of the tool.

\item \textbf{Taste}: Taste (or, the more formal term, gustation) is the ability to perceive the flavor of substances including, but not limited to, food. Humans receive tastes through sensory organs called taste buds, or gustatory calyculi, concentrated on the upper surface of the tongue. The human tongue has 100 to 150 taste receptor cells on each of its roughly ten thousand taste buds. There are five primary tastes: sweetness, bitterness, sourness, saltiness, and umami. Other tastes can be mimicked by combining these basic tastes. The recognition and awareness of umami is a relatively recent development in Western cuisine. The basic tastes contribute only partially to the sensation and flavor of food in the mouth - other factors include smell, detected by the olfactory epithelium of the nose; texture, detected through a variety of mechanoreceptors, muscle nerves, etc.; and temperature, detected by thermoreceptors. All basic tastes are classified as either appetitive or aversive, depending upon whether the things they sense are harmful or beneficial.

\item \textbf{Smell}: Smell is the process of absorbing molecules through olfactory organs. Humans absorb these molecules through the nose. These molecules diffuse through a thick layer of mucus, come into contact with one of thousands of cilia that are projected from sensory neurons, and are then absorbed into one of, 347 or so, receptors. It is this process that causes humans to understand the concept of smell from a physical standpoint. Smell is also a very interactive sense as scientists have begun to observe that olfaction comes into contact with the other sense in unexpected ways. Smell is also the most primal of the senses. It has been the discussion of being the sense that drives the most basic of human survival skills as it being the first indicator of safety or danger, friend or foe. It can be a catalyst for human behavior on a subconscious and instinctive level.

\item \textbf{Social}: Social perception is the part of perception that allows people to understand the individuals and groups of their social world, and thus an element of social cognition. People can achieve the social perception with the help of the vision, sound and touch perception, respectively. Therefore, the social perception can also cover several different sub-types listed below.
\begin{itemize}
\item \textbf{Speech}: Speech perception is the process by which spoken languages are heard, interpreted and understood. Research in speech perception seeks to understand how human listeners recognize speech sounds and use this information to understand spoken language. The sound of a word can vary widely according to words around it and the tempo of the speech, as well as the physical characteristics, accent and mood of the speaker. Listeners manage to perceive words across this wide range of different conditions. Another variation is that reverberation can make a large difference in sound between a word spoken from the far side of a room and the same word spoken up close. Experiments have shown that people automatically compensate for this effect when hearing speech.

The process of perceiving speech begins at the level of the sound within the auditory signal and the process of audition. The initial auditory signal is compared with visual information - primarily lip movement - to extract acoustic cues and phonetic information. It is possible other sensory modalities are integrated at this stage as well. This speech information can then be used for higher-level language processes, such as word recognition.

Speech perception is not necessarily uni-directional. That is, higher-level language processes connected with morphology, syntax, or semantics may interact with basic speech perception processes to aid in recognition of speech sounds. It may be the case that it is not necessary and maybe even not possible for a listener to recognize phonemes before recognizing higher units, like words for example. In one experiment, Richard M. Warren replaced one phoneme of a word with a cough-like sound. His subjects restored the missing speech sound perceptually without any difficulty and what is more, they were not able to identify accurately which phoneme had been disturbed.

\item \textbf{Face}: Facial perception refers to cognitive processes specialized for handling human faces, including perceiving the identity of an individual, and facial expressions such as emotional cues.

\item \textbf{Social Touch}: The somatosensory cortex encodes incoming sensory information from receptors all over the body. Affective touch is a type of sensory information that elicits an emotional reaction and is usually social in nature, such as a physical human touch. This type of information is actually coded differently than other sensory information. Intensity of affective touch is still encoded in the primary somatosensory cortex, but the feeling of pleasantness associated with affective touch activates the anterior cingulate cortex more than the primary somatosensory cortex. Functional magnetic resonance imaging (fMRI) data shows that increased blood oxygen level contrast (BOLD) signal in the anterior cingulate cortex as well as the prefrontal cortex is highly correlated with pleasantness scores of an affective touch. Inhibitory transcranial magnetic stimulation (TMS) of the primary somatosensory cortex inhibits the perception of affective touch intensity, but not affective touch pleasantness. Therefore, the S1 is not directly involved in processing socially affective touch pleasantness, but still plays a role in discriminating touch location and intensity.

\end{itemize}

\item \textbf{Other Types}: Other senses enable perception of body balance, acceleration, gravity, position of body parts, temperature, pain, time, and perception of internal senses such as suffocation, gag reflex, intestinal distension, fullness of rectum and urinary bladder, and sensations felt in the throat and lungs.

\end{itemize}

%--------------------------------------------------------------------------------------------------------------------
%--------------------------------------------------------------------------------------------------------------------
%--------------------------------------------------------------------------------------------------------------------

\section{Attention}

As introduced in \cite{attention}, attention is the behavioral and cognitive process of selectively concentrating on a discrete aspect of information, whether deemed subjective or objective, while ignoring other perceivable information. It is a state of arousal. It is the taking possession by the mind in clear and vivid form of one out of what seem several simultaneous objects or trains of thought. Focalization, the concentration of consciousness, is of its essence. Attention has also been described as the allocation of limited cognitive processing resources.

Attention remains a major area of investigation within education, psychology, neuroscience, cognitive neuroscience, and neuropsychology. Areas of active investigation involve determining the source of the sensory cues and signals that generate attention, the effects of these sensory cues and signals on the tuning properties of sensory neurons, and the relationship between attention and other behavioral and cognitive processes like working memory and psychological vigilance. A relatively new body of research, which expands upon earlier research within psychopathology, is investigating the diagnostic symptoms associated with traumatic brain injury and its effects on attention. Attention also varies across cultures.

The relationships between attention and consciousness are complex enough that they have warranted perennial philosophical exploration. Such exploration is both ancient and continually relevant, as it can have effects in fields ranging from mental health and the study of disorders of consciousness to artificial intelligence and its domains of research and development.

\subsection{Visual Attention}

According to \cite{attention}, in cognitive psychology there are at least two models which describe how visual attention operates. These models may be considered loosely as metaphors which are used to describe internal processes and to generate hypotheses that are falsifiable. Generally speaking, visual attention is thought to operate as a two-stage process. In the first stage, attention is distributed uniformly over the external visual scene and processing of information is performed in parallel. In the second stage, attention is concentrated to a specific area of the visual scene (i.e., it is focused), and processing is performed in a serial fashion.

The first of these models to appear in the literature is the spotlight model. The term ``spotlight'' was inspired by the work of William James, who described attention as having a focus, a margin, and a fringe. The focus is an area that extracts information from the visual scene with a high-resolution, the geometric center of which being where visual attention is directed. Surrounding the focus is the fringe of attention, which extracts information in a much more crude fashion (i.e., low-resolution). This fringe extends out to a specified area, and the cut-off is called the margin.

The second model is called the zoom-lens model and was first introduced in 1986. This model inherits all properties of the spotlight model (i.e., the focus, the fringe, and the margin), but it has the added property of changing in size. This size-change mechanism was inspired by the zoom lens one might find on a camera, and any change in size can be described by a trade-off in the efficiency of processing. The zoom-lens of attention can be described in terms of an inverse trade-off between the size of focus and the efficiency of processing: because attentional resources are assumed to be fixed, then it follows that the larger the focus is, the slower processing will be of that region of the visual scene, since this fixed resource will be distributed over a larger area. It is thought that the focus of attention can subtend a minimum of 1$^\circ$ of visual angle, however the maximum size has not yet been determined.

A significant debate emerged in the last decade of the 20th century in which Treisman's 1993 Feature Integration Theory (FIT) was compared to Duncan and Humphrey's 1989 attentional engagement theory (AET). FIT posits that ``objects are retrieved from scenes by means of selective spatial attention that picks out objects' features, forms feature maps, and integrates those features that are found at the same location into forming objects.'' Duncan and Humphrey's AET understanding of attention maintained that ``there is an initial pre-attentive parallel phase of perceptual segmentation and analysis that encompasses all of the visual items present in a scene. At this phase, descriptions of the objects in a visual scene are generated into structural units; the outcome of this parallel phase is a multiple-spatial-scale structured representation. Selective attention intervenes after this stage to select information that will be entered into visual short-term memory.'' The contrast of the two theories placed a new emphasis on the separation of visual attention tasks alone and those mediated by supplementary cognitive processes. As Rastophopoulos summarizes the debate: ``Against Treisman's FIT, which posits spatial attention as a necessary condition for detection of objects, Humphreys argues that visual elements are encoded and bound together in an initial parallel phase without focal attention, and that attention serves to select among the objects that result from this initial grouping.''

\subsection{Multitasking and Simultaneous Attention}

\subsubsection{Multitasking and Divided Attention}

Multitasking can be defined as the attempt to perform two or more tasks simultaneously; however, research shows that when multitasking, people make more mistakes or perform their tasks more slowly. Attention must be divided among all of the component tasks to perform them. In divided attention, individuals attend or give attention to multiple sources of information at once at the same time or perform more than one task.

Older research involved looking at the limits of people performing simultaneous tasks like reading stories, while listening and writing something else, or listening to two separate messages through different ears (i.e., dichotic listening). Generally, classical research into attention investigated the ability of people to learn new information when there were multiple tasks to be performed, or to probe the limits of our perception (c.f. Donald Broadbent). There is also older literature on people's performance on multiple tasks performed simultaneously, such as driving a car while tuning a radio or driving while telephoning.

The vast majority of current research on human multitasking is based on performance of doing two tasks simultaneously, usually that involves driving while performing another task, such as texting, eating, or even speaking to passengers in the vehicle, or with a friend over a cellphone. This research reveals that the human attentional system has limits for what it can process: driving performance is worse while engaged in other tasks; drivers make more mistakes, brake harder and later, get into more accidents, veer into other lanes, and/or are less aware of their surroundings when engaged in the previously discussed tasks.

There has been little difference found between speaking on a hands-free cell phone or a hand-held cell phone, which suggests that it is the strain of attentional system that causes problems, rather than what the driver is doing with his or her hands. While speaking with a passenger is as cognitively demanding as speaking with a friend over the phone, passengers are able to change the conversation based upon the needs of the driver. For example, if traffic intensifies, a passenger may stop talking to allow the driver to navigate the increasingly difficult roadway; a conversation partner over a phone would not be aware of the change in environment.

There have been multiple theories regarding divided attention. One, conceived by Kahneman, explains that there is a single pool of attentional resources that can be freely divided among multiple tasks. This model seems to be too oversimplified, however, due to the different modalities (e.g., visual, auditory, verbal) that are perceived. When the two simultaneous tasks use the same modality, such as listening to a radio station and writing a paper, it is much more difficult to concentrate on both because the tasks are likely to interfere with each other. The specific modality model was theorized by Navon and Gopher in 1979. However, more recent research using well controlled dual-task paradigms points at the importance of tasks. Specifically, in spatial visual-auditory as well as in spatial visual-tactile tasks interference of the two tasks is observed. In contrast, when one of the tasks involves object detection, no interference is observed. Thus, the multi-modal advantage in attentional resources is task dependent.

As an alternative, resource theory has been proposed as a more accurate metaphor for explaining divided attention on complex tasks. Resource theory states that as each complex task is automatized, performing that task requires less of the individual's limited-capacity attentional resources. Other variables play a part in our ability to pay attention to and concentrate on many tasks at once. These include, but are not limited to, anxiety, arousal, task difficulty, and skills.

\subsubsection{Simultaneous Attention}

Simultaneous attention is a type of attention, classified by attending to multiple events at the same time. Simultaneous attention is demonstrated by children in Indigenous communities, who learn through this type of attention to their surroundings. Simultaneous attention is present in the ways in which children of indigenous backgrounds interact both with their surroundings and with other individuals. Simultaneous attention requires focus on multiple simultaneous activities or occurrences. This differs from multitasking, which is characterized by alternating attention and focus between multiple activities, or halting one activity before switching to the next.

Simultaneous attention involves uninterrupted attention to several activities occurring at the same time. Another cultural practice that may relate to simultaneous attention strategies is coordination within a group. Indigenous heritage toddlers and caregivers in San Pedro were observed to frequently coordinate their activities with other members of a group in ways parallel to a model of simultaneous attention, whereas middle-class European-descent families in the U.S. would move back and forth between events. Research concludes that children with close ties to Indigenous American roots have a high tendency to be especially wide, keen observers. This points to a strong cultural difference in attention management.

\subsection{More Discussions on Attention}

To effectively model the attention mechanism of the human brain, several different models have been proposed, which classify the human brain attention into different categories. In this section, we will present more discussions on attention as introduced in \cite{attention}.

\subsubsection{Overt and Covert Orienting Attention}

Attention may be differentiated into ``overt'' versus ``covert'' orienting.
\begin{itemize}
\item \textbf{Overt orienting} is the act of selectively attending to an item or location over others by moving the eyes to point in that direction. Overt orienting can be directly observed in the form of eye movements. Although overt eye movements are quite common, there is a distinction that can be made between two types of eye movements; reflexive and controlled. Reflexive movements are commanded by the superior colliculus of the midbrain. These movements are fast and are activated by the sudden appearance of stimuli. In contrast, controlled eye movements are commanded by areas in the frontal lobe. These movements are slow and voluntary.

\item \textbf{Covert orienting} is the act to mentally shifting one's focus without moving one's eyes. Simply, it is changes in attention that are not attributable to overt eye movements. Covert orienting has the potential to affect the output of perceptual processes by governing attention to particular items or locations (for example, the activity of a V4 neuron whose receptive field lies on an attended stimuli will be enhanced by covert attention) but does not influence the information that is processed by the senses. Researchers often use ``filtering'' tasks to study the role of covert attention of selecting information. These tasks often require participants to observe a number of stimuli, but attend to only one.

The current view is that visual covert attention is a mechanism for quickly scanning the field of view for interesting locations. This shift in covert attention is linked to eye movement circuitry that sets up a slower saccade to that location.
\end{itemize}

There are studies that suggest the mechanisms of overt and covert orienting may not be controlled separately and independently as previously believed. Central mechanisms that may control covert orienting, such as the parietal lobe, also receive input from subcortical centers involved in overt orienting. In support of this, general theories of attention actively assume bottom-up (reflexive) processes and top-down (voluntary) processes converge on a common neural architecture, in that they control both covert and overt attentional systems. For example, if individuals attend to the right hand corner field of view, movement of the eyes in that direction may have to be actively suppressed.

\subsubsection{Exogenous and Endogenous Orienting Attention}

Orienting attention is vital and can be controlled through external (exogenous) or internal (endogenous) processes. However, comparing these two processes is challenging because external signals do not operate completely exogenously, but will only summon attention and eye movements if they are important to the subject.
\begin{itemize}
\item \textbf{Exogenous orienting} is frequently described as being under control of a stimulus. Exogenous orienting is considered to be reflexive and automatic and is caused by a sudden change in the periphery. This often results in a reflexive saccade. Since exogenous cues are typically presented in the periphery, they are referred to as peripheral cues. Exogenous orienting can even be observed when individuals are aware that the cue will not relay reliable, accurate information about where a target is going to occur. This means that the mere presence of an exogenous cue will affect the response to other stimuli that are subsequently presented in the cue's previous location.

Several studies have investigated the influence of valid and invalid cues. They concluded that valid peripheral cues benefit performance, for instance when the peripheral cues are brief flashes at the relevant location before to the onset of a visual stimulus. Posner and Cohen (1984) noted a reversal of this benefit takes place when the interval between the onset of the cue and the onset of the target is longer than about 300 ms. The phenomenon of valid cues producing longer reaction times than invalid cues is called inhibition of return.

\item \textbf{Endogenous orienting} is the intentional allocation of attentional resources to a predetermined location or space. Simply stated, endogenous orienting occurs when attention is oriented according to an observer's goals or desires, allowing the focus of attention to be manipulated by the demands of a task. In order to have an effect, endogenous cues must be processed by the observer and acted upon purposefully. These cues are frequently referred to as central cues. This is because they are typically presented at the center of a display, where an observer's eyes are likely to be fixated. Central cues, such as an arrow or digit presented at fixation, tell observers to attend to a specific location.
\end{itemize}

When examining differences between exogenous and endogenous orienting, some researchers suggest that there are four differences between the two kinds of cues:
\begin{itemize}
\item exogenous orienting is less affected by cognitive load than endogenous orienting;
\item observers are able to ignore endogenous cues but not exogenous cues;
\item exogenous cues have bigger effects than endogenous cues;
\item expectancies about cue validity and predictive value affects endogenous orienting more than exogenous orienting.
\end{itemize}

There exist both overlaps and differences in the areas of the brain that are responsible for endogenous and exogenous orientating. Another approach to this discussion has been covered under the topic heading of ``bottom-up'' versus ``top-down'' orientations to attention. Researchers of this school have described two different aspects of how the mind focuses attention to items present in the environment. The first aspect is called bottom-up processing, also known as stimulus-driven attention or exogenous attention. These describe attentional processing which is driven by the properties of the objects themselves. Some processes, such as motion or a sudden loud noise, can attract our attention in a pre-conscious, or non-volitional way. We attend to them whether we want to or not. These aspects of attention are thought to involve parietal and temporal cortices, as well as the brainstem.

The second aspect is called top-down processing, also known as goal-driven, endogenous attention, attentional control or executive attention. This aspect of our attentional orienting is under the control of the person wh

\subsubsection{Perceptual and Cognitive Attention}

Meanwhile, the Perceptual load theory states that there are two mechanisms regarding selective attention: perceptual and cognitive. 
\begin{itemize}
\item The \textbf{perceptual attention} considers the subject's ability to perceive or ignore stimuli, both task-related and non task-related. Studies show that if there are many stimuli present (especially if they are task-related), it is much easier to ignore the non-task related stimuli, but if there are few stimuli the mind will perceive the irrelevant stimuli as well as the relevant. 

\item The \textbf{cognitive attention} refers to the actual processing of the stimuli. Studies regarding this showed that the ability to process stimuli decreased with age, meaning that younger people were able to perceive more stimuli and fully process them, but were likely to process both relevant and irrelevant information, while older people could process fewer stimuli, but usually processed only relevant information.
\end{itemize}

Some people can process multiple stimuli, e.g. trained morse code operators have been able to copy 100\% of a message while carrying on a meaningful conversation. This relies on the reflexive response due to ``overlearning'' the skill of morse code reception/detection/transcription so that it is an autonomous function requiring no specific attention to perform.

\subsubsection{Clinical Model on Attention}

Attention is best described as the sustained focus of cognitive resources on information while filtering or ignoring extraneous information. Attention is a very basic function that often is a precursor to all other neurological/cognitive functions. As is frequently the case, clinical models of attention differ from investigation models. One of the most used models for the evaluation of attention in patients with very different neurologic pathologies is the model of Sohlberg and Mateer. This hierarchic model is based in the recovering of attention processes of brain damage patients after coma. Five different kinds of activities of growing difficulty are described in the model; connecting with the activities those patients could do as their recovering process advanced.
\begin{itemize}
\item \textbf{Focused attention}: The ability to respond discretely to specific visual, auditory or tactile stimuli.
\item \textbf{Sustained attention (vigilance and concentration)}: The ability to maintain a consistent behavioral response during continuous and repetitive activity.
\item \textbf{Selective attention}: The ability to maintain a behavioral or cognitive set in the face of distracting or competing stimuli. Therefore, it incorporates the notion of ``freedom from distractibility.''
\item \textbf{Alternating attention}: The ability of mental flexibility that allows individuals to shift their focus of attention and move between tasks having different cognitive requirements.
\item \textbf{Divided attention}: This refers to the ability to respond simultaneously to multiple tasks or multiple task demands.
\end{itemize}

This model has been shown to be very useful in evaluating attention in very different pathologies, correlates strongly with daily difficulties and is especially helpful in designing stimulation programs such as attention process training, a rehabilitation program for neurological patients of the same authors.

\begin{itemize}
\item \textbf{Mindfulness}: Mindfulness has been conceptualized as a clinical model of attention. Mindfulness practices are clinical interventions that emphasize training attention functions.
\end{itemize}

%--------------------------------------------------------------------------------------------------------------------
%--------------------------------------------------------------------------------------------------------------------
%--------------------------------------------------------------------------------------------------------------------

\section{Memory}

According to \cite{memory_intro}, memory is our ability to encode, store, retain and subsequently recall information and past experiences in the human brain. It can be thought of in general terms as the use of past experience to affect or influence current behavior. Memory is the sum total of what we remember, and gives us the capability to learn and adapt from previous experiences as well as to build relationships. It is the ability to remember past experiences, and the power or process of recalling to mind previously learned facts, experiences, impressions, skills and habits. It is the store of things learned and retained from our activity or experience, as evidenced by modification of structure or behavior, or by recall and recognition.

In more physiological or neurological terms, memory is, at its simplest, a set of encoded neural connections in the brain. It is the re-creation or reconstruction of past experiences by the synchronous firing of neurons that were involved in the original experience. As we will see, though, because of the way in which memory is encoded, it is perhaps better thought of as a kind of collage or jigsaw puzzle, rather than in the traditional manner as a collection of recordings or pictures or video clips, stored as discrete wholes. Our memories are not stored in our brains like books on library shelves, but are actually on-the-fly reconstructions from elements scattered throughout various areas of our brains.

As introduced in \cite{memory_intro_2}, it seems that our memory is located not in one particular place in the brain, but is instead a brain-wide process in which several different areas of the brain act in conjunction with one another (sometimes referred to as distributed processing). For example, the simple act of riding a bike is actively and seamlessly reconstructed by the brain from many different areas: the memory of how to operate the bike comes from one area, the memory of how to get from here to the end of the block comes from another, the memory of biking safety rules from another, and that nervous feeling when a car veers dangerously close comes from still another. Each element of a memory (sights, sounds, words, emotions) is encoded in the same part of the brain that originally created that fragment (visual cortex, motor cortex, language area, etc), and recall of a memory effectively reactivates the neural patterns generated during the original encoding. Thus, a better image might be that of a complex web, in which the threads symbolize the various elements of a memory, that join at nodes or intersection points to form a whole rounded memory of a person, object or event. This kind of distributed memory ensures that even if part of the brain is damaged, some parts of an experience may still remain. Neurologists are only beginning to understand how the parts are reassembled into a coherent whole.

Memory is related to but distinct from learning, which is the process by which we acquire knowledge of the world and modify our subsequent behavior. During learning, neurons that fire together to produce a particular experience are altered so that they have a tendency to fire together again. For example, we learn a new language by studying it, but we then speak it by using our memory to retrieve the words that we have learned. Thus, memory depends on learning because it lets us store and retrieve learned information. But learning also depends to some extent on memory, in that the knowledge stored in our memory provides the framework to which new knowledge is linked by association and inference. This ability of humans to call on past memories in order to imagine the future and to plan future courses of action is a hugely advantageous attribute in our survival and development as a species.

Neither is memory a single unitary process but there are different types of memory. Our short term and long-term memories are encoded and stored in different ways and in different parts of the brain, for reasons that we are only beginning to guess at. Years of case studies of patients suffering from accidents and brain-related diseases and other disorders (especially in elderly persons) have begun to indicate some of the complexities of the memory processes, and great strides have been made in neuroscience and cognitive psychology, but many of the exact mechanisms involved remain elusive.

\subsection{Types of Memory}

\begin{figure}[t]
    \centering
    \includegraphics[width=0.8\textwidth]{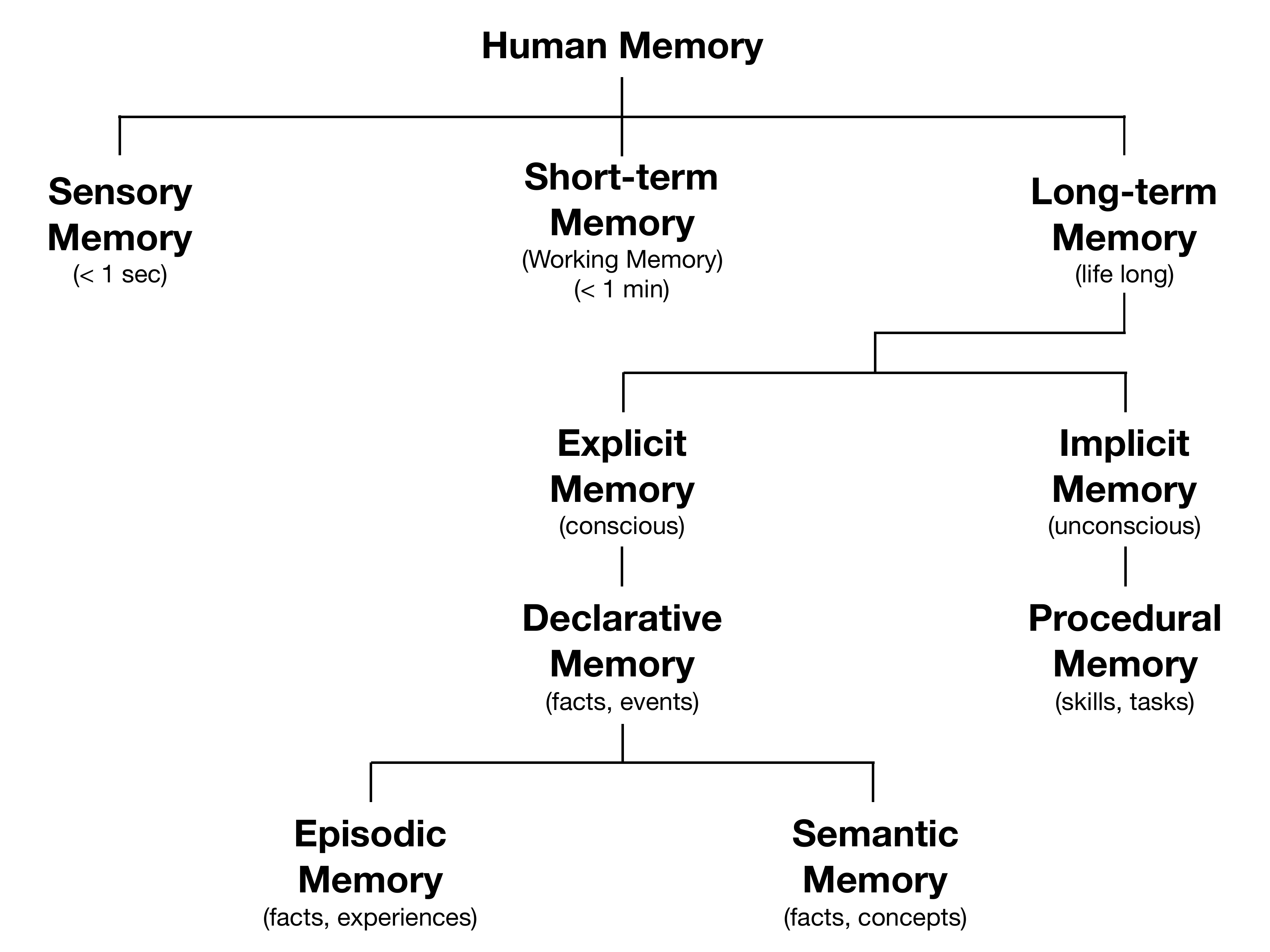}
    \caption{Types of Human Memory \cite{memory_type}.}
    \label{fig:memory_type}
\end{figure}

What we usually think of as ``memory'' in day-to-day usage is actually long-term memory, but there are also important short-term and sensory memory processes, which must be worked through before a long-term memory can be established. The different types of memory each have their own particular mode of operation, but they all cooperate in the process of memorization, and can be seen as three necessary steps in forming a lasting memory. As illustrated in Figure~\ref{fig:memory_type}, we provide a complete diagram of different types of human memory as well as their hierarchical relationships. 

%------------------------
%------------------------

\subsubsection{Sensory Memory}

As introduced in \cite{memory_sensory}, sensory memory is the shortest-term element of memory. It is the ability to retain impressions of sensory information after the original stimuli have ended. It acts as a kind of buffer for stimuli received through the five senses of sight, hearing, smell, taste and touch, which are retained accurately, but very briefly. For example, the ability to look at something and remember what it looked like with just a second of observation is an example of sensory memory.

The stimuli detected by our senses can be either deliberately ignored, in which case they disappear almost instantaneously, or perceived, in which case they enter our sensory memory. This does not require any conscious attention and, indeed, is usually considered to be totally outside of conscious control. The brain is designed to only process information that will be useful at a later date, and to allow the rest to pass by unnoted. As information is perceived, it is therefore stored in sensory memory automatically and unbidden. Unlike other types of memory, the sensory memory cannot be prolonged via rehearsal.

Sensory memory is an ultra-short-term memory and decays or degrades very quickly, typically in the region of 200 - 500 milliseconds (1/5 - 1/2 second) after the perception of an item, and certainly less than a second (although echoic memory is now thought to last a little longer, up to perhaps three or four seconds). Indeed, it lasts for such a short time that it is often considered part of the process of perception, but it nevertheless represents an essential step for storing information in short-term memory.

The sensory memory for visual stimuli is sometimes known as the iconic memory, the memory for aural stimuli is known as the echoic memory, and that for touch as the haptic memory. Smell may actually be even more closely linked to memory than the other senses, possibly because the olfactory bulb and olfactory cortex (where smell sensations are processed) are physically very close - separated by just 2 or 3 synapses - to the hippocampus and amygdala (which are involved in memory processes). Thus, smells may be more quickly and more strongly associated with memories and their associated emotions than the other senses, and memories of a smell may persist for longer, even without constant re-consolidation.

Experiments by George Sperling in the early 1960s involving the flashing of a grid of letters for a very short period of time (50 milliseconds) suggest that the upper limit of sensory memory (as distinct from short-term memory) is approximately 12 items, although participants often reported that they seemed to ``see'' more than they could actually report.

Information is passed from the sensory memory into short-term memory via the process of attention (the cognitive process of selectively concentrating on one aspect of the environment while ignoring other things), which effectively filters the stimuli to only those which are of interest at any given time.

\subsubsection{Short-Term Working Memory}

According to \cite{memory_short}, short-term memory acts as a kind of ``scratch-pad'' for temporary recall of the information which is being processed at any point in time, and has been referred to as ``the brain's Post-it note''. It can be thought of as the ability to remember and process information at the same time. It holds a small amount of information (typically around 7 items or even less) in mind in an active, readily-available state for a short period of time (typically from 10 to 15 seconds, or sometimes up to a minute).

For example, in order to understand this sentence, the beginning of the sentence needs to be held in mind while the rest is read, a task which is carried out by the short-term memory. Other common examples of short-term memory in action are the holding on to a piece of information temporarily in order to complete a task (e.g. ``carrying over'' a number in a subtraction sum, or remembering a persuasive argument until another person finishes talking), and simultaneous translation (where the interpreter must store information in one language while orally translating it into another). What is actually held in short-term memory, though, is not complete concepts, but rather links or pointers (such as words, for example) which the brain can flesh out from it's other accumulated knowledge.

The central executive part of the prefrontal cortex at the front of the brain appears to play a fundamental role in short-term and working memory. It both serves as a temporary store for short-term memory, where information is kept available while it is needed for current reasoning processes, but it also ``calls up'' information from elsewhere in the brain. The central executive controls two neural loops, one for visual data (which activates areas near the visual cortex of the brain and acts as a visual scratch pad), and one for language (the ``phonological loop'', which uses Broca's area as a kind of ``inner voice'' that repeats word sounds to keep them in mind). These two scratch pads temporarily hold data until it is erased by the next job. Although the prefrontal cortex is not the only part of the brain involved - it must also cooperate with other parts of the cortex from which it extracts information for brief periods - it is the most important, and Carlyle Jacobsen reported, as early as 1935, that damage to the prefrontal cortex in primates caused short-term memory deficits.

The short-term memory has a limited capacity, which can be readily illustrated by the simple expedient of trying to remember a list of random items (without allowing repetition or reinforcement) and seeing when errors begin to creep in. The often-cited experiments by George Miller in 1956 suggest that the number of objects an average human can hold in working memory (known as memory span) is between 5 and 9 (7 $\pm$ 2, which Miller described as the ``magical number'', and which is sometimes referred to as Miller's Law). However, although this may be approximately true for a population of college students, for example, memory span varies widely with populations tested, and modern estimates are typically lower, of the order of just 4 or 5 items.

``Chunking'' of information can lead to an increase in the short-term memory capacity. Chunking is the organization of material into shorter meaningful groups to make them more manageable. For example, a hyphenated phone number, split into groups of 3 or 4 digits, tends to be easier to remember than a single long number. Experiments by Herbert Simon have shown that the ideal size for chunking of letters and numbers, whether meaningful or not, is three. However, meaningful groups may be longer (such as four numbers that make up a date within a longer list of numbers, for example). With chunking, each chunk represents just one of the 5-9 items that can be stored in short-term memory, thus extending the total number of items that can be held.

It is usually assumed that the short-term memory spontaneously decays over time, typically in the region of 10-15 seconds, but items may be retained for up to a minute, depending on the content. However, it can be extended by repetition or rehearsal (either by reading items out loud, or by mental simulation), so that the information re-enters the short-term store and is retained for a further period. When several elements (such as digits, words or pictures) are held in short-term memory simultaneously, they effectively compete with each other for recall. New content, therefore, gradually pushes out older content (known as displacement), unless the older content is actively protected against interference by rehearsal or by directing attention to it. Any outside interference tends to cause disturbances in short-term memory retention, and for this reason people often feel a distinct desire to complete the tasks held in short-term memory as soon as possible.

Typically, information is transferred from the short-term or working memory to the long-term memory within just a few seconds, although the exact mechanisms by which this transfer takes place, and whether all or only some memories are retained permanently, remain controversial topics among experts. Richard Schiffrin, in particular, is well known for his work in the 1960s suggesting that all memories automatically pass from a short-term to a long-term store after a short time (known as the modal or multi-store or Atkinson-Schiffrin model).

However, this is disputed, and it now seems increasingly likely that some kind of vetting or editing procedure takes place. Some researchers (e.g. Eugen Tarnow) have proposed that there is no real distinction between short-term and long-term memory at all, and certainly it is difficult to demarcate a clear boundary between them. However, the evidence of patients with some kinds of anterograde amnesia, and experiments on the way distraction affect the short-term recall of lists, suggest that there are in fact two more or less separate systems.

\subsubsection{Long-Term Memory}

Long-term memory is, obviously enough, intended for storage of information over a long period of time. Despite our everyday impressions of forgetting, it seems likely that long-term memory actually decays very little over time, and can store a seemingly unlimited amount of information almost indefinitely. Indeed, there is some debate as to whether we actually ever ``forget'' anything at all, or whether it just becomes increasingly difficult to access or retrieve certain items from memory.

Short-term memories can become long-term memory through the process of consolidation, involving rehearsal and meaningful association. Unlike short-term memory (which relies mostly on an acoustic, and to a lesser extent a visual, code for storing information), long-term memory encodes information for storage semantically (i.e. based on meaning and association). However, there is also some evidence that long-term memory does also encode to some extent by sound. For example, when we cannot quite remember a word but it is ``on the tip of the tongue'', this is usually based on the sound of a word, not its meaning.

Physiologically, the establishment of long-term memory involves a process of physical changes in the structure of neurons (or nerve cells) in the brain, a process known as long-term potentiation, although there is still much that is not completely understood about the process. At its simplest, whenever something is learned, circuits of neurons in the brain, known as neural networks, are created, altered or strengthened. These neural circuits are composed of a number of neurons that communicate with one another through special junctions called synapses. Through a process involving the creation of new proteins within the body of neurons, and the electrochemical transfer of neurotransmitters across synapse gaps to receptors, the communicative strength of certain circuits of neurons in the brain is reinforced. With repeated use, the efficiency of these synapse connections increases, facilitating the passage of nerve impulses along particular neural circuits, which may involve many connections to the visual cortex, the auditory cortex, the associative regions of the cortex, etc.

This process differs both structurally and functionally from the creation of working or short-term memory. Although the short-term memory is supported by transient patterns of neuronal communication in the regions of the frontal, prefrontal and parietal lobes of the brain, long-term memories are maintained by more stable and permanent changes in neural connections widely spread throughout the brain. The hippocampus area of the brain essentially acts as a kind of temporary transit point for long-term memories, and is not itself used to store information. However, it is essential to the consolidation of information from short-term to long-term memory, and is thought to be involved in changing neural connections for a period of three months or more after the initial learning.

Unlike with short-term memory, forgetting occurs in long-term memory when the formerly strengthened synaptic connections among the neurons in a neural network become weakened, or when the activation of a new network is superimposed over an older one, thus causing interference in the older memory. Over the years, several different types of long-term memory have been distinguished, including explicit and implicit memory, declarative and procedural memory (with a further sub-division of declarative memory into episodic and semantic memory) and retrospective and prospective memory.\\

\noindent \textbf{Explicit and Implicit Memory (or Declarative and Procedural Memory)}

As introduced in \cite{memory_explicit_implicit}, long-term memory is often divided into two further main types: explicit (or declarative) memory and implicit (or procedural) memory.

\begin{itemize}
\item Declarative memory (``knowing what'') is memory of facts and events, and refers to those memories that can be consciously recalled (or ``declared''). It is sometimes called explicit memory, since it consists of information that is explicitly stored and retrieved, although it is more properly a subset of explicit memory. Declarative memory can be further sub-divided into episodic memory and semantic memory.

\item Procedural memory (``knowing how'') is the unconscious memory of skills and how to do things, particularly the use of objects or movements of the body, such as tying a shoelace, playing a guitar or riding a bike. These memories are typically acquired through repetition and practice, and are composed of automatic sensorimotor behaviors that are so deeply embedded that we are no longer aware of them. Once learned, these ``body memories'' allow us to carry out ordinary motor actions more or less automatically. Procedural memory is sometimes referred to as implicit memory, because previous experiences aid in the performance of a task without explicit and conscious awareness of these previous experiences, although it is more properly a subset of implicit memory.
\end{itemize}

These different types of long-term memory are stored in different regions of the brain and undergo quite different processes. Declarative memories are encoded by the hippocampus, entorhinal cortex and perirhinal cortex (all within the medial temporal lobe of the brain), but are consolidated and stored in the temporal cortex and elsewhere. Procedural memories, on the other hand, do not appear to involve the hippocampus at all, and are encoded and stored by the cerebellum, putamen, caudate nucleus and the motor cortex, all of which are involved in motor control. Learned skills such as riding a bike are stored in the putamen; instinctive actions such as grooming are stored in the caudate nucleus; and the cerebellum is involved with timing and coordination of body skills. Thus, without the medial temporal lobe (the structure that includes the hippocampus), a person is still able to form new procedural memories (such as playing the piano, for example), but cannot remember the events during which they happened or were learned.\\

\noindent \textbf{Episodic and Semantic Memory}

Declarative memory can be further sub-divided into episodic memory and semantic memory.

\begin{itemize}
\item Episodic memory represents our memory of experiences and specific events in time in a serial form, from which we can reconstruct the actual events that took place at any given point in our lives. It is the memory of autobiographical events (times, places, associated emotions and other contextual knowledge) that can be explicitly stated. Individuals tend to see themselves as actors in these events, and the emotional charge and the entire context surrounding an event is usually part of the memory, not just the bare facts of the event itself.

\item Semantic memory, on the other hand, is a more structured record of facts, meanings, concepts and knowledge about the external world that we have acquired. It refers to general factual knowledge, shared with others and independent of personal experience and of the spatial/temporal context in which it was acquired. Semantic memories may once have had a personal context, but now stand alone as simple knowledge. It therefore includes such things as types of food, capital cities, social customs, functions of objects, vocabulary, understanding of mathematics, etc. Much of semantic memory is abstract and relational and is associated with the meaning of verbal symbols.
\end{itemize}

The semantic memory is generally derived from the episodic memory, in that we learn new facts or concepts from our experiences, and the episodic memory is considered to support and underpin semantic memory. A gradual transition from episodic to semantic memory can take place, in which episodic memory reduces its sensitivity and association to particular events, so that the information can be generalized as semantic memory.

Both episodic memory and semantic memory require a similar encoding process. However, semantic memory mainly activates the frontal and temporal cortexes, whereas episodic memory activity is concentrated in the hippocampus, at least initially. Once processed in the hippocampus, episodic memories are then consolidated and stored in the neocortex. The memories of the different elements of a particular event are distributed in the various visual, olfactory and auditory areas of the brain, but they are all connected together by the hippocampus to form an episode, rather than remaining a collection of separate memories.\\

\noindent \textbf{Retrospective and Prospective Memory}

An important alternative classification of long-term memory used by some researchers is based on the temporal direction of the memories.

\begin{itemize}
\item Retrospective memory is where the content to be remembered (people, words, events, etc) is in the past, i.e. the recollection of past episodes. It includes semantic, episodic and autobiographical memory, and declarative memory in general, although it can be either explicit or implicit.

\item Prospective memory is where the content is to be remembered in the future, and may be defined as ``remembering to remember'' or remembering to perform an intended action. It may be either event-based or time-based, often triggered by a cue, such as going to the doctor (action) at 4pm (cue), or remembering to post a letter (action) after seeing a mailbox (cue).
\end{itemize}

Clearly, though, retrospective and prospective memory are not entirely independent entities, and certain aspects of retrospective memory are usually required for prospective memory. Thus, there have been case studies where an impaired retrospective memory has caused a definite impact on prospective memory. However, there have also been studies where patients with an impaired prospective memory had an intact retrospective memory, suggesting that to some extent the two types of memory involve separate processes.

\subsection{Memory Process}

\begin{figure}[t]
    \centering
    \includegraphics[width=0.8\textwidth]{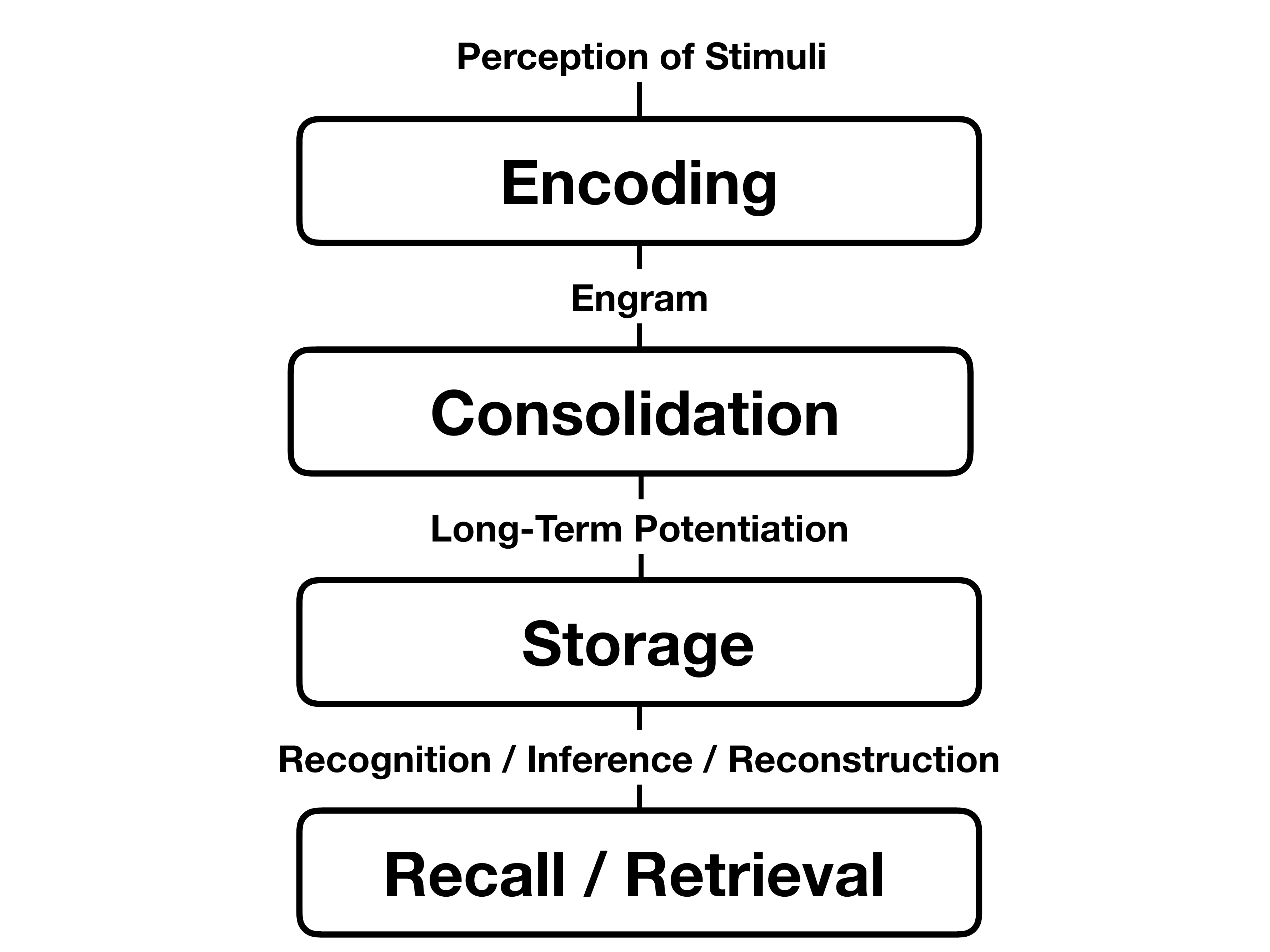}
    \caption{Human Memory Formation Process \cite{memory_process}.}
    \label{fig:memory_process}
\end{figure}

As introduced in \cite{memory_process}, memory is the ability to encode, store and recall information. The three main processes involved in human memory are therefore encoding, storage and recall (retrieval). Additionally, the process of memory consolidation (which can be considered to be either part of the encoding process or the storage process) is treated here as a separate process in its own right. As shown in Figure~\ref{fig:memory_process}, the relationships among these processes and the memory formation are clearly illustrated.

%------------------------
%------------------------

\subsubsection{Memory Encoding}

According to \cite{memory_encoding}, encoding is the crucial first step to creating a new memory. It allows the perceived item of interest to be converted into a construct that can be stored within the brain, and then recalled later from short-term or long-term memory.

Encoding is a biological event beginning with perception through the senses. The process of laying down a memory begins with attention (regulated by the thalamus and the frontal lobe), in which a memorable event causes neurons to fire more frequently, making the experience more intense and increasing the likelihood that the event is encoded as a memory. Emotion tends to increase attention, and the emotional element of an event is processed on an unconscious pathway in the brain leading to the amygdala. Only then are the actual sensations derived from an event processed.

The perceived sensations are decoded in the various sensory areas of the cortex, and then combined in the brain's hippocampus into one single experience. The hippocampus is then responsible for analyzing these inputs and ultimately deciding if they will be committed to long-term memory. It acts as a kind of sorting centre where the new sensations are compared and associated with previously recorded ones. The various threads of information are then stored in various different parts of the brain, although the exact way in which these pieces are identified and recalled later remains largely unknown. The key role that the hippocampus plays in memory encoding has been highlighted by examples of individuals who have had their hippocampus damaged or removed and can no longer create new memories (see Anterograde Amnesia). It is also one of the few areas of the brain where completely new neurons can grow.

Although the exact mechanism is not completely understood, encoding occurs on different levels, the first step being the formation of short-term memory from the ultra-short term sensory memory, followed by the conversion to a long-term memory by a process of memory consolidation. The process begins with the creation of a memory trace or engram in response to the external stimuli. An engram is a hypothetical biophysical or biochemical change in the neurons of the brain, hypothetical in the respect that no-one has ever actually seen, or even proved the existence of, such a construct.

An organ called the hippocampus, deep within the medial temporal lobe of the brain, receives connections from the primary sensory areas of the cortex, as well as from associative areas and the rhinal and entorhinal cortexes. While these anterograde connections converge at the hippocampus, other retrograde pathways emerge from it, returning to the primary cortexes. A neural network of cortical synapses effectively records the various associations which are linked to the individual memory.

There are three or four main types of encoding:
\begin{itemize}
\item Acoustic encoding is the processing and encoding of sound, words and other auditory input for storage and later retrieval. This is aided by the concept of the phonological loop, which allows input within our echoic memory to be sub-vocally rehearsed in order to facilitate remembering.

\item Visual encoding is the process of encoding images and visual sensory information. Visual sensory information is temporarily stored within the iconic memory before being encoded into long-term storage. The amygdala (within the medial temporal lobe of the brain which has a primary role in the processing of emotional reactions) fulfills an important role in visual encoding, as it accepts visual input in addition to input from other systems and encodes the positive or negative values of conditioned stimuli.

\item Tactile encoding is the encoding of how something feels, normally through the sense of touch. Physiologically, neurons in the primary somatosensory cortex of the brain react to vibrotactile stimuli caused by the feel of an object.

\item Semantic encoding is the process of encoding sensory input that has particular meaning or can be applied to a particular context, rather than deriving from a particular sense.
\end{itemize}

It is believed that, in general, encoding for short-term memory storage in the brain relies primarily on acoustic encoding, while encoding for long-term storage is more reliant (although not exclusively) on semantic encoding.

Human memory is fundamentally associative, meaning that a new piece of information is remembered better if it can be associated with previously acquired knowledge that is already firmly anchored in memory. The more personally meaningful the association, the more effective the encoding and consolidation. Elaborate processing that emphasizes meaning and associations that are familiar tends to leads to improved recall. On the other hand, information that a person finds difficult to understand cannot be readily associated with already acquired knowledge, and so will usually be poorly remembered, and may even be remembered in a distorted form due to the effort to comprehend its meaning and associations. For example, given a list of words like ``thread'', ``sewing'', ``haystack'', ``sharp'', ``point'', ``syringe'', ``pin'', ``pierce'', ``injection'' and ``knitting'', people often also (incorrectly) remember the word ``needle'' through a process of association.

Because of the associative nature of memory, encoding can be improved by a strategy of organization of memory called elaboration, in which new pieces of information are associated with other information already recorded in long-term memory, thus incorporating them into a broader, coherent narrative which is already familiar. An example of this kind of elaboration is the use of mnemonics, which are verbal, visual or auditory associations with other, easy-to-remember constructs, which can then be related back to the data that is to be remembered. Rhymes, acronymns, acrostics and codes can all be used in this way. Common examples are ``Roy G. Biv'' to remember the order of the colours of the rainbow, or ``Every Good Boy Deserves Favour'' for the musical notes on the lines of the treble clef, which most people find easier to remember than the original list of colours or letters. When we use mnemonic devices, we are effectively passing facts through the hippocampus several times, so that it can keep strengthening the associations, and therefore improve the likelihood of subsequent memory recall.

In the same way, associating words with images is another commonly used mnemonic device, providing two alternative methods of remembering, and creating additional associations in the mind. Taking this to a higher level, another method of improving memory encoding and consolidation is the use of a so-called memory palace (also known as the method of loci), a mnemonic techniques that relies on memorized spatial relationships to establish, order and recollect other memories. The method is to assign objects or facts to different rooms in an imaginary house or palace, so that recall of the facts can be cued by mentally ``walking though'' the palace until it is found. Many top memorizers today use the memory palace method to a greater or lesser degree. Similar techniques involve placing the items at different landmarks on a favourite hike or trip (known as the journey method), or weaving them into a story.

\subsubsection{Memory Consolidation}

As introduced in \cite{memory_consolidation}, consolidation is the processes of stabilizing a memory trace after the initial acquisition. It may perhaps be thought of part of the process of encoding or of storage, or it may be considered as a memory process in its own right. It is usually considered to consist of two specific processes, synaptic consolidation (which occurs within the first few hours after learning or encoding) and system consolidation (where hippocampus-dependent memories become independent of the hippocampus over a period of weeks to years).

Neurologically, the process of consolidation utilizes a phenomenon called long-term potentiation, which allows a synapse to increase in strength as increasing numbers of signals are transmitted between the two neurons. Potentiation is the process by which synchronous firing of neurons makes those neurons more inclined to fire together in the future. Long-term potentiation occurs when the same group of neurons fire together so often that they become permanently sensitized to each other. As new experiences accumulate, the brain creates more and more connections and pathways, and may ``re-wire'' itself by re-routing connections and re-arranging its organization.

As such a neuronal pathway, or neural network, is traversed over and over again, an enduring pattern is engraved and neural messages are more likely to flow along such familiar paths of least resistance. This process is achieved by the production of new proteins to rebuild the synapses in the new shape, without which the memory remains fragile and easily eroded with time. For example, if a piece of music is played over and over, the repeated firing of certain synapses in a certain order in your brain makes it easier to repeat this firing later on, with the result that the musician becomes better at playing the music, and can play it faster, with fewer mistakes.

In this way, the brain organizes and reorganizes itself in response to experiences, creating new memories prompted by experience, education or training. The ability of the connection, or synapse, between two neurons to change in strength, and for lasting changes to occur in the efficiency of synaptic transmission, is known as synaptic plasticity or neural plasticity, and it is one of the important neurochemical foundations of memory and learning.

It should be remembered that each neuron makes thousands of connections with other neurons, and memories and neural connections are mutually interconnected in extremely complex ways. Unlike the functioning of a computer, each memory is embedded in many connections, and each connection is involved in several memories. Thus, multiple memories may be encoded within a single neural network by different patterns of synaptic connections. Conversely, a single memory may involve simultaneously activating several different groups of neurons in completely different parts of the brain.

The inverse of long-term potentiation, known as long-term depression, can also take place, whereby the neural networks involved in erroneous movements are inhibited by the silencing of their synaptic connections. This can occur in the cerebellum, which is located towards the back of the brain, in order to correct our motor procedures when learning how to perform a task (procedural memory), but also in the synapses of the cortex, the hippocampus, the striatum and other memory-related structures.

Contrary to long-term potentiation, which is triggered by high-frequency stimulation of the synapses, long-term depression is produced by nerve impulses reaching the synapses at very low frequencies, leading them to undergo the reverse transformation from long-term potentiation, and, instead of becoming more efficient, the synaptic connections are weakened. It is still not clear whether long-term depression contributes directly to the storage of memories in some way, or whether it simply makes us forget the traces of some things learned long ago so that new things can be learned.

Sleep (particularly slow-wave, or deep, sleep, during the first few hours) is also thought to be important in improving the consolidation of information in memory, and activation patterns in the sleeping brain, which mirror those recorded during the learning of tasks from the previous day, suggest that new memories may be solidified through such reactivation and rehearsal.

Memory re-consolidation is the process of previously consolidated memories being recalled and then actively consolidated all over again, in order to maintain, strengthen and modify memories that are already stored in the long-term memory. Several retrievals of memory (either naturally through reflection, or through deliberate recall) may be needed for long-term memories to last for many years, depending on the depth of the initial processing. However, these individual retrievals can take place at increasing intervals, in accordance with the principle of spaced repetition (this is familiar to us in the way that ``cramming'' the night before an exam is not as effective as studying at intervals over a much longer span of time).

The very act of re-consolidation, though, may change the initial memory. As a particular memory trace is reactivated, the strengths of the neural connections may change, the memory may become associated with new emotional or environmental conditions or subsequently acquired knowledge, expectations rather than actual events may become incorporated into the memory, etc.

\subsubsection{Memory Storage}

According to \cite{processes_storage}, memory storage is the more or less passive process of retaining information in the brain, whether in the sensory memory, the short-term memory or the more permanent long-term memory. Each of these different stages of human memory function as a sort of filter that helps to protect us from the flood of information that confront us on a daily basis, avoiding an overload of information and helping to keep us sane. The more the information is repeated or used, the more likely it is to be retained in long-term memory (which is why, for example, studying helps people to perform better on tests). This process of consolidation, the stabilizing of a memory trace after its initial acquisition, is treated in more detail in a separate section.

Since the early neurological work of Karl Lashley and Wilder Penfield in the 1950s and 1960s, it has become clear that long-term memories are not stored in just one part of the brain, but are widely distributed throughout the cortex. After consolidation, long-term memories are stored throughout the brain as groups of neurons that are primed to fire together in the same pattern that created the original experience, and each component of a memory is stored in the brain area that initiated it (e.g. groups of neurons in the visual cortex store a sight, neurons in the amygdala store the associated emotion, etc). Indeed, it seems that they may even be encoded redundantly, several times, in various parts of the cortex, so that, if one engram (or memory trace) is wiped out, there are duplicates, or alternative pathways, elsewhere, through which the memory may still be retrieved.

Therefore, contrary to the popular notion, memories are not stored in our brains like books on library shelves, but must be actively reconstructed from elements scattered throughout various areas of the brain by the encoding process. Memory storage is therefore an ongoing process of reclassification resulting from continuous changes in our neural pathways, and parallel processing of information in our brains.

The indications are that, in the absence of disorders due to trauma or neurological disease, the human brain has the capacity to store almost unlimited amounts of information indefinitely. Forgetting, therefore, is more likely to be result from incorrectly or incompletely encoded memories, and/or problems with the recall/retrieval process. It is a common experience that we may try to remember something one time and fail, but then remember that same item later. The information is therefore clearly still there in storage, but there may have been some kind of a mismatch between retrieval cues and the original encoding of the information. ``Lost'' memories recalled with the aid of psychotherapy or hypnosis are other examples supporting this idea, although it is difficult to be sure that such memories are real and not implanted by the treatment.

Having said that, though, it seems unlikely that, as Richard Schiffrin and others have claimed, ALL memories are stored somewhere in the brain, and that it is only in the retrieval process that irrelevant details are ``fast-forwarded'' over or expurgated. It seems more likely that the memories which are stored are in some way edited and sorted, and that some of the more peripheral details are never stored.

Forgetting, then, is perhaps better thought of as the temporary or permanent inability to retrieve a piece of information or a memory that had previously been recorded in the brain. Forgetting typically follows a logarithmic curve, so that information loss is quite rapid at the start, but becomes slower as time goes on. In particular, information that has been learned very well (e.g. names, facts, foreign-language vocabulary, etc), will usually be very resistant to forgetting, especially after the first three years. Unlike amnesia, forgetting is usually regarded as a normal phenomenon involving specific pieces of content, rather than relatively broad categories of memories or even entire segments of memory.

Theorists disagree over exactly what becomes of material that is forgotten. Some hold that long-term memories do actually decay and disappear completely over time; others hold that the memory trace remains intact as long as we live, but the bonds or cues that allow us to retrieve the trace become broken, due to changes in the organization of the neural network, new experiences, etc, in the same way as a misplaced book in a library is ``lost'' even though it still exists somewhere in the library.

Increasing forgetfulness is a normal part of the aging process, as the neurons in aging brains lose their connections and start to die off, and, ultimately the brain shrinks and becomes less effective. The hippocampus, which as we have seen is crucial for memory and learning, is one of the first areas of the brain to deteriorate with age. Recent studies in mice involving infusions of blood from young mice into older mice have shown that the old mice that received young blood showed a significant burst of brain cell growth in the hippocampus region (and vice versa), leading to speculation that young blood might represent the antidote to senile forgetfulness (and other ravages of old age). Similar studies on humans with Alzheimers disease are currently in progress.

Interestingly, it appears NOT to be possible to deliberately delete memories at will, which can have negative consequences, for example if we experience traumatic events we would actually prefer to forget. In fact, such memories tend to be imprinted even more strongly than normal due to their emotional content, although recent research involving the use of beta blockers suggests that it may be possible to tone down the emotional aspects of such memories, even if the memories themselves cannot be erased. The way this works is that the act of recalling stored memories makes them ``malleable'' once more, as they were during the initial encoding phase, and their re-storage can then be blocked by drugs which inhibit the proteins that enable the emotional memory to be re-saved.

\subsubsection{Memory Recall/Retrieval}

Recall or retrieval of memory \cite{processes_recall} refers to the subsequent re-accessing of events or information from the past, which have been previously encoded and stored in the brain. In common parlance, it is known as remembering. During recall, the brain ``replays'' a pattern of neural activity that was originally generated in response to a particular event, echoing the brain's perception of the real event. In fact, there is no real solid distinction between the act of remembering and the act of thinking.

These replays are not quite identical to the original, though - otherwise we would not know the difference between the genuine experience and the memory - but are mixed with an awareness of the current situation. One corollary of this is that memories are not frozen in time, and new information and suggestions may become incorporated into old memories over time. Thus, remembering can be thought of as an act of creative reimagination.

Because of the way memories are encoded and stored, memory recall is effectively an on-the-fly reconstruction of elements scattered throughout various areas of our brains. Memories are not stored in our brains like books on library shelves, or even as a collection of self-contained recordings or pictures or video clips, but may be better thought of as a kind of collage or a jigsaw puzzle, involving different elements stored in disparate parts of the brain linked together by associations and neural networks. Memory retrieval therefore requires re-visiting the nerve pathways the brain formed when encoding the memory, and the strength of those pathways determines how quickly the memory can be recalled. Recall effectively returns a memory from long-term storage to short-term or working memory, where it can be accessed, in a kind of mirror image of the encoding process. It is then re-stored back in long-term memory, thus re-consolidating and strengthening it.

The efficiency of human memory recall is astounding. Most of what we remember is by direct retrieval, where items of information are linked directly a question or cue, rather than by the kind of sequential scan a computer might use (which would require a systematic search through the entire contents of memory until a match is found). Other memories are retrieved quickly and efficiently by hierarchical inference, where a specific question is linked to a class or subset of information about which certain facts are known. Also, the brain is usually able to determine in advance whether there is any point in searching memory for a particular fact (e.g. it instantly recognizes a question like ``What is Socrates' telephone number?'' as absurd in that no search could ever produce an answer).

There are two main methods of accessing memory: recognition and recall. 
\begin{itemize}
\item Recognition is the association of an event or physical object with one previously experienced or encountered, and involves a process of comparison of information with memory, e.g. recognizing a known face, true/false or multiple choice questions, etc. Recognition is a largely unconscious process, and the brain even has a dedicated face-recognition area, which passes information directly through the limbic areas to generate a sense of familiarity, before linking up with the cortical path, where data about the person's movements and intentions are processed. 
\item Recall involves remembering a fact, event or object that is not currently physically present (in the sense of retrieving a representation, mental image or concept), and requires the direct uncovering of information from memory, e.g. remembering the name of a recognized person, fill-in the blank questions, etc.
\end{itemize}

Recognition is usually considered to be ``superior'' to recall (in the sense of being more effective), in that it requires just a single process rather than two processes. Recognition requires only a simple familiarity decision, whereas a full recall of an item from memory requires a two-stage process (indeed, this is often referred to as the two-stage theory of memory) in which the search and retrieval of candidate items from memory is followed by a familiarity decision where the correct information is chosen from the candidates retrieved. Thus, recall involves actively reconstructing the information and requires the activation of all the neurons involved in the memory in question, whereas recognition only requires a relatively simple decision as to whether one thing among others has been encountered before. Sometimes, however, even if a part of an object initially activates only a part of the neural network concerned, recognition may then suffice to activate the entire network.

There are three main types of recall:
\begin{itemize}
\item Free recall is the process in which a person is given a list of items to remember and then is asked to recall them in any order (hence the name ``free''). This type of recall often displays evidence of either the primacy effect (when the person recalls items presented at the beginning of the list earlier and more often) or the recency effect (when the person recalls items presented at the end of the list earlier and more often), and also of the contiguity effect (the marked tendency for items from neighboring positions in the list to be recalled successively).

\item Cued recall is the process in which a person is given a list of items to remember and is then tested with the use of cues or guides. When cues are provided to a person, they tend to remember items on the list that they did not originally recall without a cue, and which were thought to be lost to memory. This can also take the form of stimulus-response recall, as when words, pictures and numbers are presented together in a pair, and the resulting associations between the two items cues the recall of the second item in the pair.

\item Serial recall refers to our ability to recall items or events in the order in which they occurred, whether chronological events in our autobiographical memories, or the order of the different parts of a sentence (or phonemes in a word) in order to make sense of them. Serial recall in long-term memory appears to differ from serial recall in short-term memory, in that a sequence in long-term memory is represented in memory as a whole, rather than as a series of discrete items. Testing of serial recall by psychologists have yielded several general rules:
\begin{itemize}
\item more recent events are more easily remembered in order (especially with auditory stimuli);
\item recall decreases as the length of the list or sequence increases;
\item there is a tendency to remember the correct items, but in the wrong order;
\item where errors are made, there is a tendency to respond with an item that resembles the original item in some way (e.g. ``dog'' instead of ``fog'', or perhaps an item physically close to the original item);
\item repetition errors do occur, but they are relatively rare;
\item if an item is recalled earlier in the list than it should be, the missed item tends to be inserted immediately after it;
\item if an item from a previous trial is recalled in a current trial, it is likely to be recalled at its position from the original trial.
\end{itemize}
\end{itemize}

If we assume that the ``purpose'' of human memory is to use past events to guide future actions, then keeping a perfect and complete record of every past event is not necessarily a useful or efficient way of achieving this. So, in most people, some specific memories may be given up or converted into general knowledge (i.e. converted from episodic to semantic memories) as part of the ongoing recall/re-consolidation process, so that that we are able to generalize from experience.

%--------------------------------------------------------------------------------------------------------------------
%--------------------------------------------------------------------------------------------------------------------
%--------------------------------------------------------------------------------------------------------------------

%\section{Learning}
%\section{Consciousness}

\section{Summary}

Based on the basic neural units introduced in \cite{zhang2019basic} and the overall structure described in \cite{zhang2019secrets}, the brain neural network can perform many complex functions and activities. As the last of the three series tutorial articles on the brain, in this paper, we provide an introduction to the brain cognitive functions. More specifically, we have covered the perception, attention and memory in this paper. This paper together with the previous two are will be updated accordingly as we observe new development in neuroscience.

\newpage

\vskip 0.2in
\bibliographystyle{plain}
\bibliography{reference}

\begin{thebibliography}{10}

\bibitem{attention}
Attention.
\newblock \url{https://en.wikipedia.org/wiki/Attention}.
\newblock [Online; accessed 27-May-2019].

\bibitem{cognition}
Cognition.
\newblock \url{https://en.wikipedia.org/wiki/Cognition}.
\newblock [Online; accessed 27-May-2019].

\bibitem{memory_explicit_implicit}
Declarative (explicit) \& procedural (implicit) memory.
\newblock \url{http://www.human-memory.net/types_declarative.html}.
\newblock [Online; accessed 27-May-2019].

\bibitem{memory_consolidation}
Memory consolidation.
\newblock \url{http://www.human-memory.net/processes_consolidation.html}.
\newblock [Online; accessed 27-May-2019].

\bibitem{memory_encoding}
Memory encoding.
\newblock \url{http://www.human-memory.net/processes_encoding.html}.
\newblock [Online; accessed 27-May-2019].

\bibitem{memory_process}
Memory processes.
\newblock \url{http://www.human-memory.net/processes.html}.
\newblock [Online; accessed 27-May-2019].

\bibitem{processes_recall}
Memory recall/retrieval.
\newblock \url{http://www.human-memory.net/processes_recall.html}.
\newblock [Online; accessed 27-May-2019].

\bibitem{processes_storage}
Memory storage.
\newblock \url{http://www.human-memory.net/processes_storage.html}.
\newblock [Online; accessed 27-May-2019].

\bibitem{perception}
Perception.
\newblock \url{https://en.wikipedia.org/wiki/Perception}.
\newblock [Online; accessed 27-May-2019].

\bibitem{memory_sensory}
Sensory memory.
\newblock \url{http://www.human-memory.net/types_sensory.html}.
\newblock [Online; accessed 27-May-2019].

\bibitem{memory_short}
Short-term (working) memory.
\newblock \url{http://www.human-memory.net/types_short.html}.
\newblock [Online; accessed 27-May-2019].

\bibitem{memory_intro_2}
The human memory.
\newblock \url{http://www.human-memory.net/index.html}.
\newblock [Online; accessed 27-May-2019].

\bibitem{memory_type}
Types of memory.
\newblock \url{http://www.human-memory.net/types.html}.
\newblock [Online; accessed 27-May-2019].

\bibitem{memory_intro}
What is memory?
\newblock \url{http://www.human-memory.net/intro_what.html}.
\newblock [Online; accessed 27-May-2019].

\bibitem{cognition_table}
Pascale Michelon.
\newblock What are cognitive abilities and skills, and how to boost them?
\newblock
  \url{https://sharpbrains.com/blog/2006/12/18/what-are-cognitive-abilities/}.
\newblock [Online; accessed 27-May-2019].

\bibitem{zhang2019basic}
Jiawei Zhang.
\newblock Basic neural units of the brain: Neurons, synapses and action
  potential, 2019.

\bibitem{zhang2019secrets}
Jiawei Zhang.
\newblock Secrets of the brain: An introduction to the brain anatomical
  structure and biological function, 2019.

\end{thebibliography}

\end{document}